# 合肥光源储存环逐束团 3 维位置测量系统介绍及初步研究

卢平，唐雷雷，孙葆根，吴芳芳，吴睿喆，王季刚，周泽然

（中国科学技术大学 国家同步辐射实验室，合肥 230029）

**摘要**　　合肥光源储存环束流位置检测器BPM的 4 路电极信号直接接入国产 12bit分辨率、10Gsps采样率、2GHz带宽的示波器，采集程序运行在Zstack架构下的云主机上，每次触发读取 500μs的一组波形，提取出 45 个束团 2266 圈束团质心的X（水平位置）、Y（垂直位置）、Z（纵向相位）信息，逐束团的X、Y的分辨率约为 5μm，Z的分辨率约为 0.5ps，在线运行更新周期 7 秒左右。得到的逐束团 3 维位置信息做频谱分析可得到运行常态时每个束团的 3 维工作点，对纽扣和条带电极信号的分析可得到激励时横向尺寸四极振荡的谱峰。

**关键词**：合肥光源；逐束团；3 维参数；四极振荡；BBQ；EPICS；Zstack；caLab；LabView

## 一、系统概况

观测储存环中束流的逐束团 3 维运动情况是非常有意义的事[1][2]，现在仪器技术的发展，特别是国产仪器的各项指标达到了甚至好于各光源在线常用的国际巨头的仪器性能指标，并且稳定性也达到了常年在线运行的要求，使得基于国产示波器测量逐束团 3 维运动信息成为可能，本论文工作使用鼎阳公司的型号为 SDS6204 H12 Pro[3]的示波器，在 Zstack[4]虚拟架构的 Win10 云主机上使用 LabView 开发环境编写采集和处理程序，程序中通过 caLab 把 EPICS 的 PV 变量写入到 Centos 系统的 IOC，之后通过 OPI 获取结果并展示。

表 1 给出了合肥光源储存环参数表[5]。图 1 是合肥光源储存环束测件的布局图，本系统测量的是 7 号位置处的条带纽扣组合 BPM，图 2 是逐束团 3 维位置测量系统框图。

表 1　　合肥光源储存环参数表

| | |
|---|---|
| 束流能量 GeV | 0.8 |
| 高频频率[MHz] | 204.030MHz |
| 谐波数 | 45 |
| 辐射能量损失 [keV] | 16.73 |
| 自然束流发射度 [ nm·rad] | <40 |
| 注入流强 [mA] | 400mA |
| 自然能散(rms) | 0.00047 |
| 束流寿命 [hours] | >5hours |
| 自然束团长度 (mm) | 14.8 (约 50 ps ) |

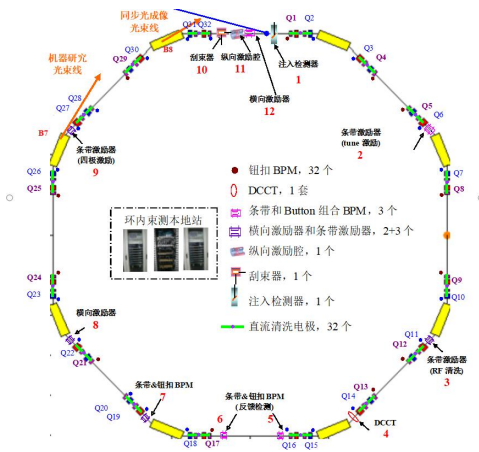
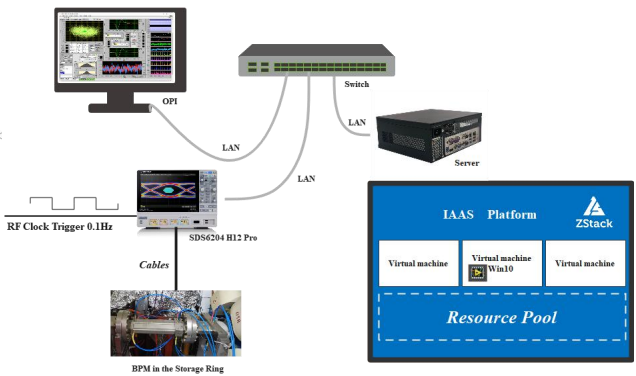

图 1  合肥光源储存环束测件的布局图　　　　图 2  逐束团 3 维位置测量系统框图

为了方便对比，使用两台鼎阳该款示波器，一台接条带 BPM，一台接纽扣 BPM，通过一个分频器，把时序系统 1Hz 的触发信号分频成 0.1Hz 作为示波器的触发，图 3 是系统实物图。图 4 中有 3 组 BPM，左边两组在同一段真空室上，左边那组是条带 BPM，右边那组为纽扣 BPM，分别接入两个示波器；最右边四极铁旁的那组纽扣 BPM 是 COD 系统中的 IVU1，接入 Libera B+处理器[6]。

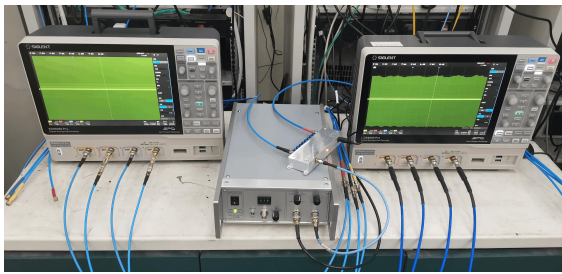
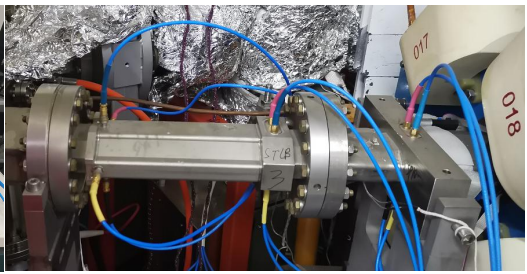

图 3  系统实物图　　　　图 4  3 组 BPM

每次触发，采集程序获取 500μs 长，每个通道 5M 个数据点的四组波形，去掉头尾非整圈的波形点，共包含 2266 整圈的逐束团信息，通过过零点插值法提取出逐束团的时间信息、通过每个束团极值处及左右各一点共 3 点找抛物线极值点[7]或简单积分、平方和积分后开方等方法（程序中还有各方法融合功能，可选择和融合或者积融合）提取幅度信息，之后通过差比和换算出逐束团水平和垂直位置(X、Y)，此外利用 DCCT 的流强数据标定换算出逐束团的流强。除了发布逐束团的信息，程序中还拆分得到每个束团的逐圈信息，并做频谱分析后通过 PV 发布，可得到每个束团的横向和纵向工作点信息。程序中还可选择是否重采样、陷波滤波等预处理功能，并包含了 SVD 分解选择各成分的功能。程序处理一组波形并发布的周期为 7 秒，主要瓶颈在读取波形，4 组 5M 个点的波形共 40MB 的数据，要花 7 秒才能从示波器全部读到云主机里，平均速度 5.7MB/s，为了两台示波器获取同一段时间的波形分析结果并对比，通过一个分频器触发，触发周期为 10 秒。程序界面如图 5 所示。

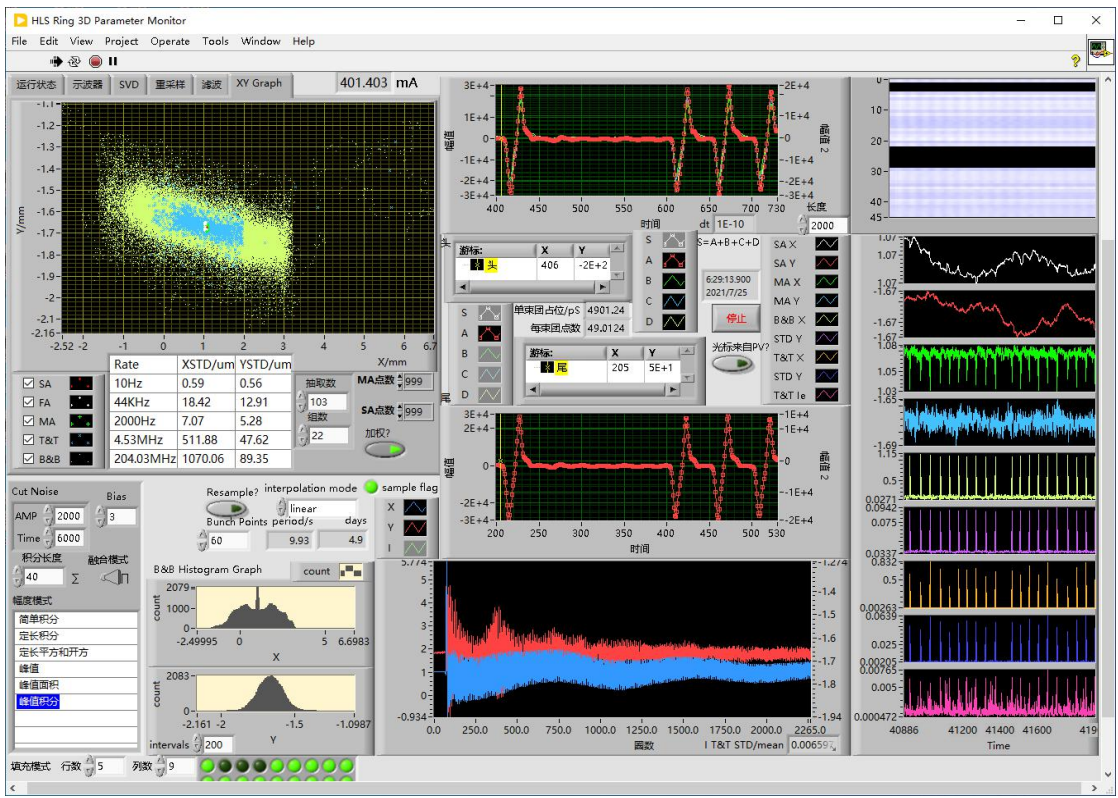

图 5    程序界面

**二、测量结果**

得到一组波形的逐束团 X、Y、Z 的信息后，可以把 2266 圈的 45 个束团的 3 维位置通过 3 视图展示出来，或者要看某个逐束团信息的话，可以把束团编号和圈数做为二维图的坐标，通过不同颜色展示该信息（波纹图）。图 6 给出了波纹图可以很直观的看到纵向振荡的情况。

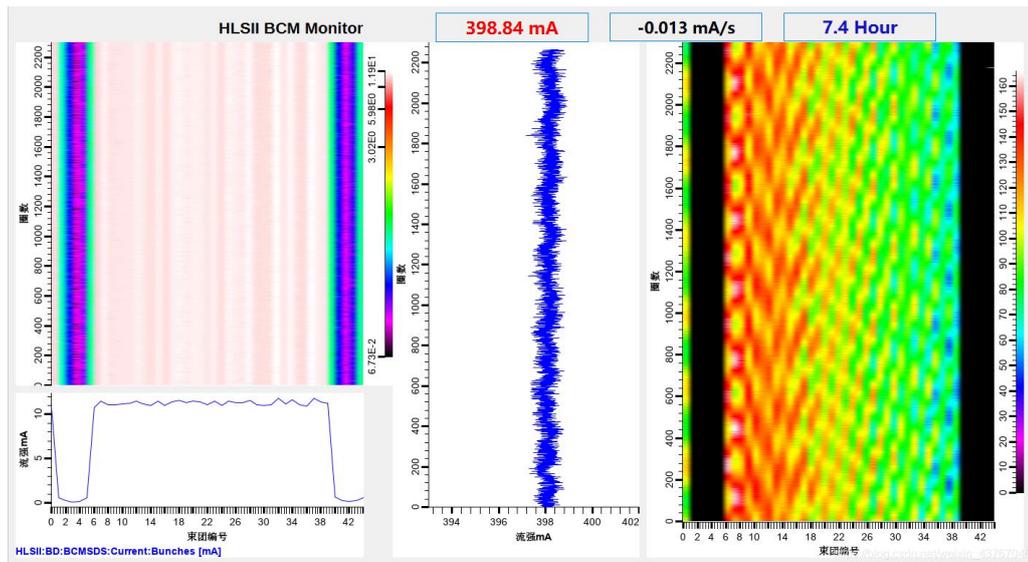

图 6    波纹图显示的纵向振荡的情况

图 6 左边是逐束团流强的二维图，下面的曲线是每个束团 2266 圈的平均流强；中间的曲线是 45 个束团的流强求和后的逐圈流强曲线；右边的是纵向时间的二维图。

图 7 给出了注入时条带和纽扣 X、Y、I 的波纹图，下面的曲线是频谱，可看到工作点的谱峰。

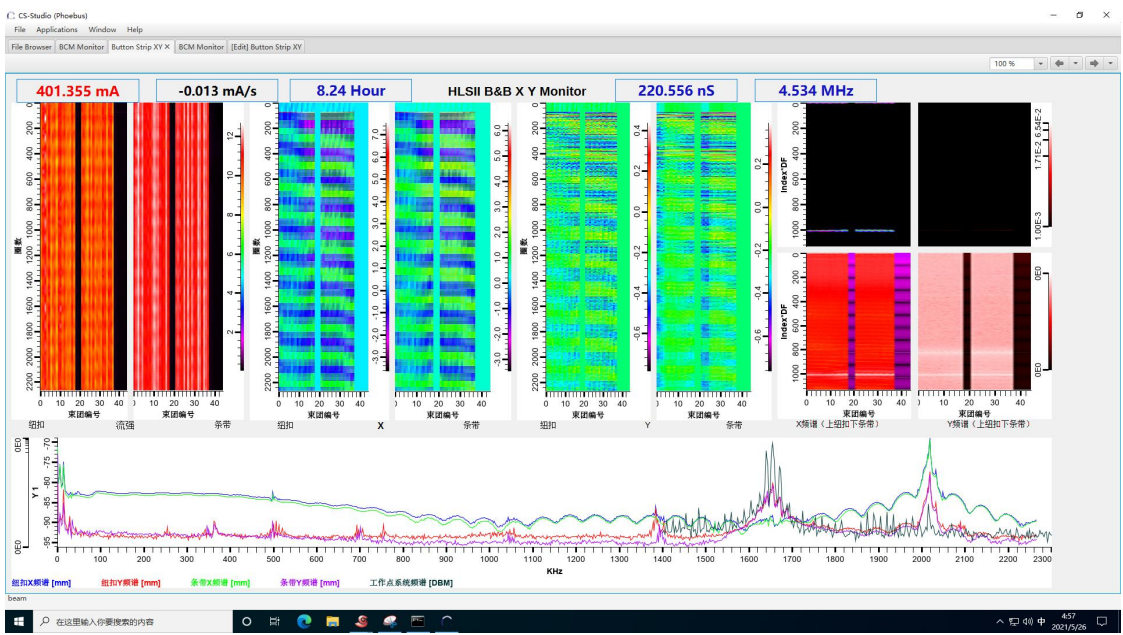

图 7 注入时条带和纽扣 X、Y、I 的波纹图

从三视图可以很直观的看到每个束团的运动情况，图 8 为正常运行状态，图 9 为注入状态，三视图左边那组为条带 BPM，右边为纽扣 BPM，不同颜色代表不同的束团，其中最上面的红色点群为 45 个束团 3 维位置平均后的逐圈坐标在三视图中的分布。

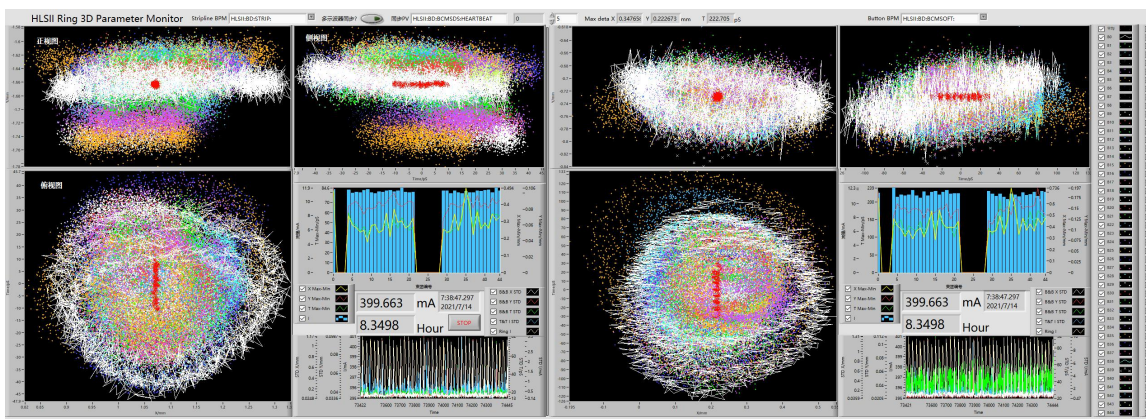

图 8  正常运行状态下的三视图

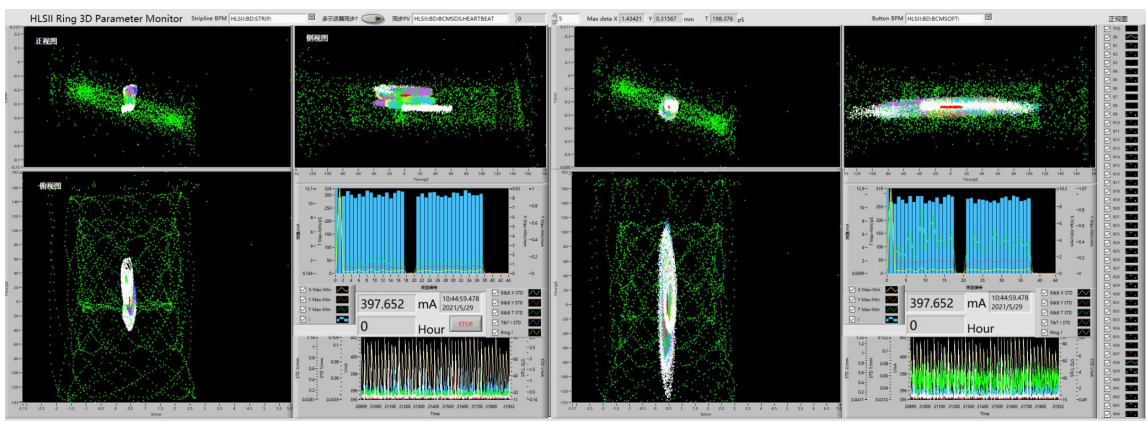

图 9 注入状态下的三视图

### 三、误差分析

通过观察 3 视图发现束团因为纵向振荡耦合到 X 方向而使俯视图有明显的画圈的现象，正视图有时也会发现有这样的现象，画圈的低频部分去除后通过计算标准差 STD 可以分析测量误差，图 10 的 3 幅图是取某个束团的测量值减去画圈的低频部分的分布图，得到的 X，Y，Z 的 STD 值，分别约为 3.8μm、4.2μm、0.4ps，各个束团以及不同组波形的数据相差不大，向上取到整分别为 5μm、5μm、0.5ps。

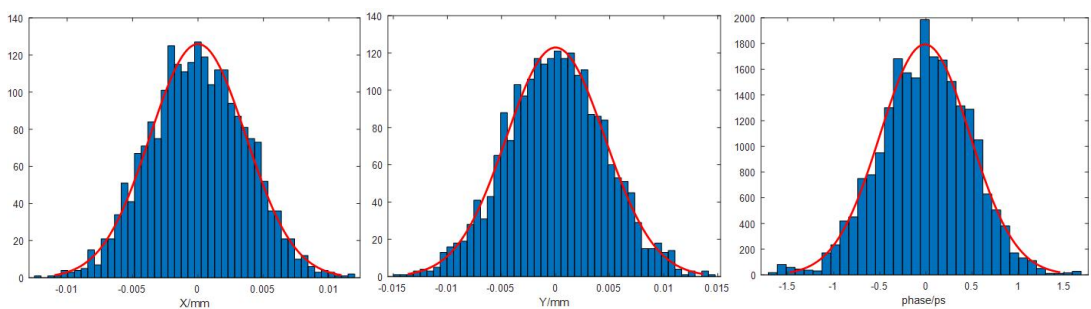

图 10　逐束团 3 维位置的测量值减去画圈的低频部分的分布图

一组波形（500μs）逐束团的 X、Y 可以逐次平均得到逐圈、FA（快速获取 44kHz）、MA（中速获取 2kHz）的数据，由于两台示波器为了对比，处理周期为 10s，如果每 500μs 波形平均得到一个 X、Y 的话，等效为一个 2KHz 的信号（相当于 MA）10 秒抽取一个点，为了和 Libera B+测量的 SA（10Hz）结果对比，需要把从 2KHz 信号抽取的 200 个点做平均再对比。

图 11 是程序在线工作时获取某组波形计算经过逐次平均得到的标准差 STD 结果，图 12 为 Libera B+测量 COD 系统在线状态显示。

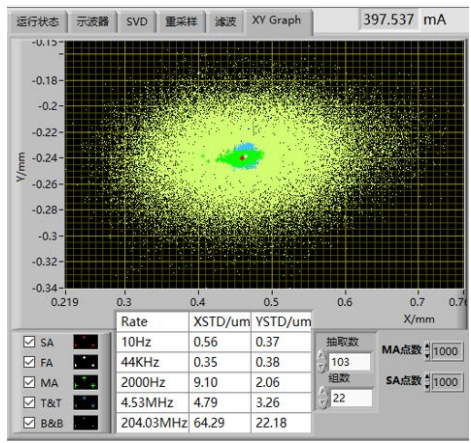 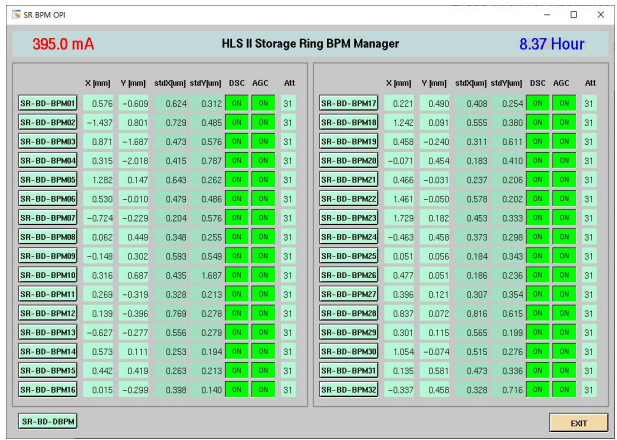

图 11 逐次平均得到的 STD 结果　　图 12　Libera B+测量 COD 系统在线状态

Libera B+测量的 COD 系统的 32 个 BPM 的 X、Y 的 SA 数据的 STD 都在 1μm 以内，示波器滚动平均得到的 SA 的 STD 和其相当，一组波形的逐圈、FA 数据好于 MA 和 SA，主要因为逐圈和 FA 的 STD 是在一组波形 500μs 内的数据统计，而 MA 和 SA 是不同组波形的数据更长时间的统计。

图 13 给出了示波器的 SA 和 Libera B+的 SA 连续 2.5 天监测对比。图 13(a)为水平方向 X，绿色曲线是 Libera B+ 测的 IVU1，蓝色和橙色分别是纽扣和条带，红色是 DCCT；图 13(b)为垂直方向 Y，蓝色曲线是 Libera B+ 测的 IVU1，橙色和绿色分别是纽扣和条带，红色是 DCCT。

从长期对比结果看，Libera B+一直稳定在 3、4μm 范围内，如果忽略流强变化大的地方，示波器的 X 在 20μm、Y 在 10μm 范围内漂动，这应该和示波器温控的措施，以及长期工作没有交联开关等消除通道不一致性的手段有关，由此还产生结果的流强依赖性（特别是 X 要严重些）。

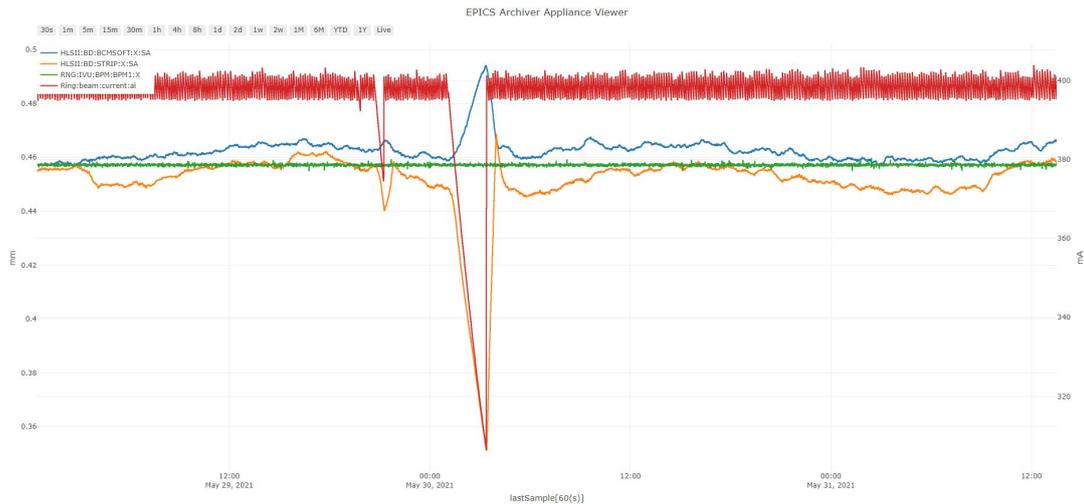

(a) 水平方向 X

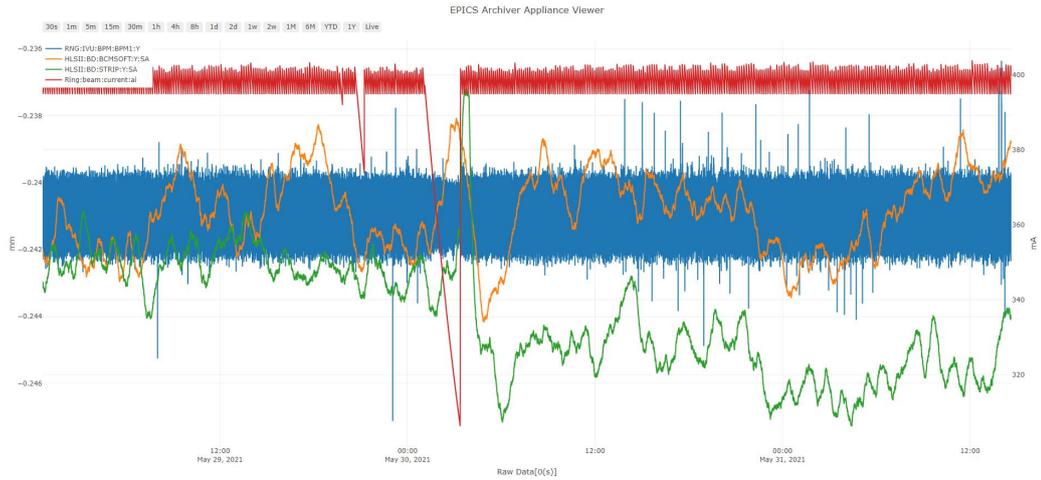

(b) 垂直方向 Y

图 13    示波器的 SA 和 Libera B+的 SA 连续 2.5 天监测对比

图 14 给出了 COD 闭轨反馈系统关闭，任由轨道漂动的情况下，示波器的 MA 和 Libera B+的 SA 连续 24 小时监测对比，红色曲线是 Libera B+测的 IVU1，橙色和绿色分别是纽扣和条带，蓝色是 DCCT。图 14(a)为水平方向 X，图 14(b)为垂直方向 Y。

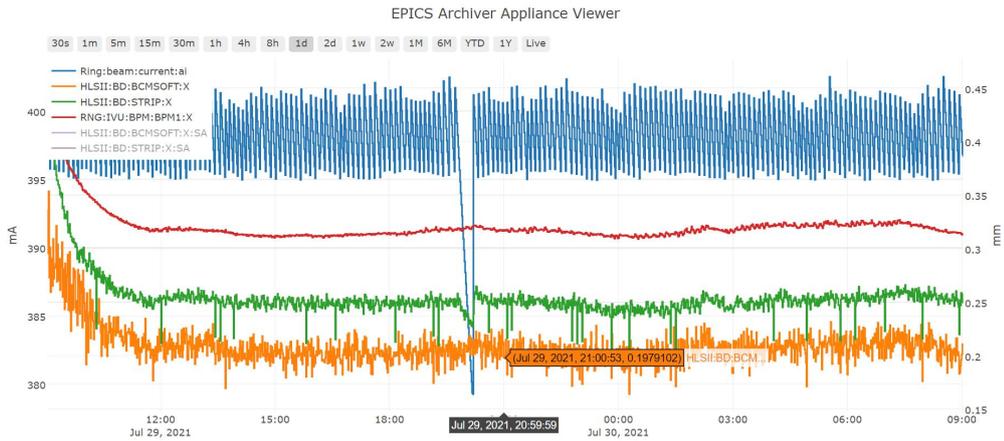

(a)    水平方向 X

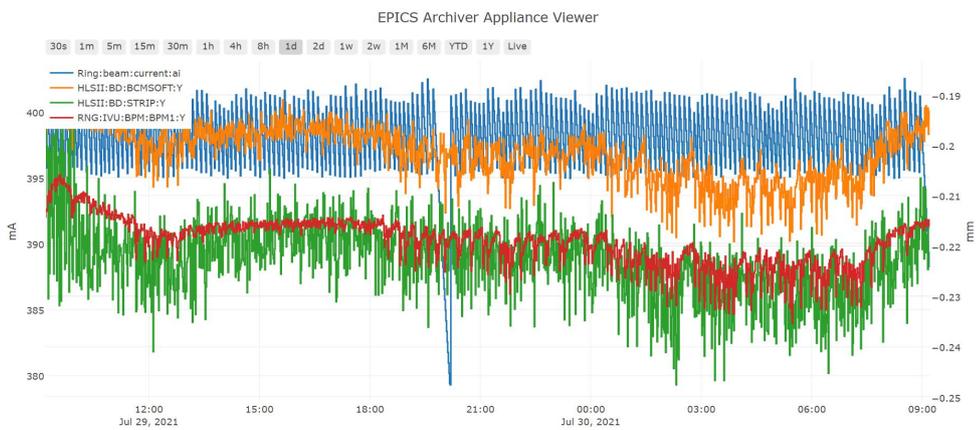

(b)    垂直方向 Y

图 14  不加闭轨反馈情况下，示波器的 MA 和 Libera B+的 SA 连续 24 小时监测对比

## 四、初步分析结果
### 1. 工作点

图 15 为无束流时 X、Y 的频谱。由图 15 可见，无束流时 X、Y 的频谱曲线在约 350KHz 及其多倍频处有峰，这些是噪声在幅度提取计算 X、Y 时产生的，查看有束流时的谱线可以忽略这些地方的谱峰。

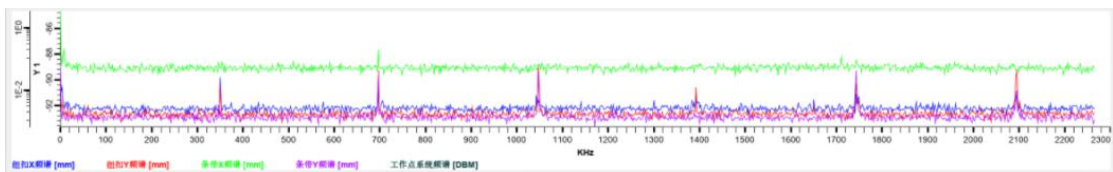

图 15 无束流时 I、X、Y 的二维图和频谱

图 16 为运行常态时的频谱。观察图 16 的频谱，上面的 Turn By Turn 的频谱图是 45 个束团各自的频谱之后求和后的曲线，合肥光源的回旋频率为 $f_0$=4.534MHz，所以频谱能看到的范围为 0~$f_0$/2=2.267MHz；下面的 Bunch By Bunch 是逐束团 X、Y 的频谱，能看到的范围为 0~45×$f_0$/2=102MHz。从图 16 可以看到工作点位置的谱峰。

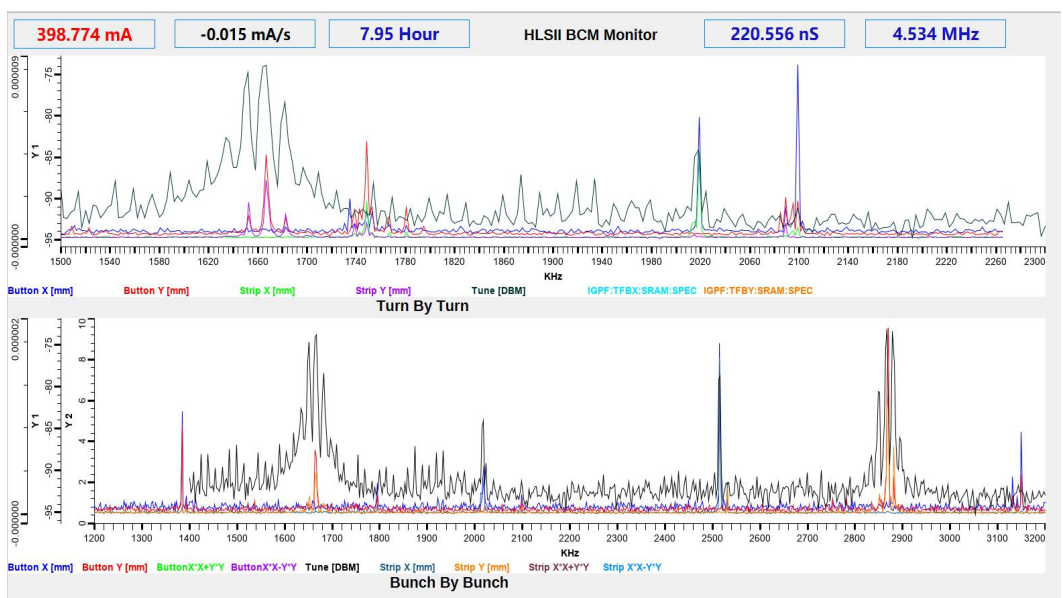

图 16 运行常态时的频谱

从频谱曲线上看，信噪比和分辨率普遍好于 BBQ 系统的谱线，除了和 BBQ 谱线对应的工作点处的谱峰外，其他地方的谱峰都正好是图 15 显示的那些噪声峰位。通过逐束团数据，可以分析出每个束团的工作点。

图 17 分别为 45 个束团 X、Y、Z 的工作点频率直方图，水平工作点，平均的频率 2.0094MHz，对应小数部分为 0.4432；垂直工作点，平均的频率 1.6732MHz，对应小数部分为 0.3690；纵向工作点，平均的频率为 14.44kHz，对应工作点为 0.0032。

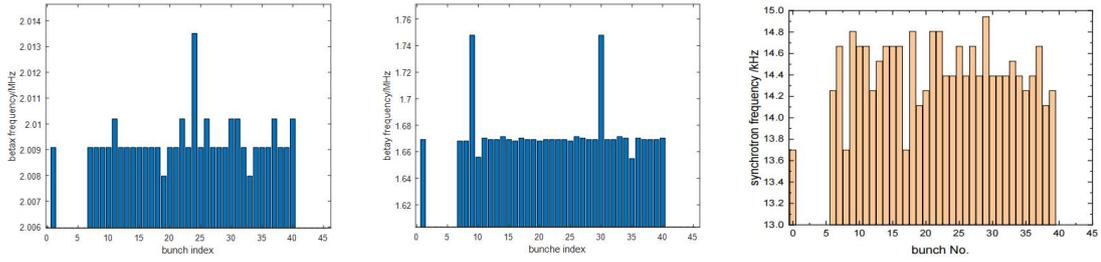

图 17　45 个束团 X、Y、Z 的工作点频率直方图

## 2. 四极振荡[8]

### 2.1 束流横向四极振荡和二极振荡的关系

储存环中束流横向二极振荡是束流中粒子集体 Beta 振荡的表现，振荡频率即为工作点对应的频率，表示为 $(n \pm \Delta \nu_{x,y})f_0$，其中 $n$ 为整数，$\Delta \nu_{x,y}$ 分别为水平和垂直工作点的小数部分，$f_0$ 为回旋频率。束流横向 Beta 振荡是储存环横向聚焦 Lattice 结构的反映，而横向四极振荡反映的是束流本身的特性。为了描述横向四极振荡与二极振荡之间的关系，将粒子的类正弦振荡简化为正弦振荡，如图 18 所示，考虑单束团中水平方向开始时相距最远的两个粒子的运动。

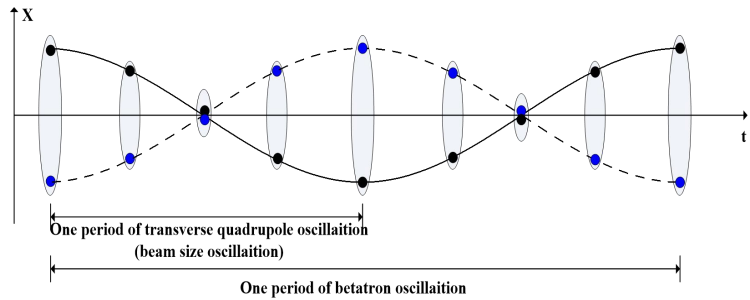

图 18　束流横向四极振荡和二极振荡的关系示意图

束团横向四极振荡对应的是束团横向尺寸的振荡，如上图所示，当束团水平方向尺寸振荡了一个周期时，该束团中水平方向开始时相距最远的两个粒子都只运动了半个 Beta 振荡周期。粒子 Beta 振荡频率可以表示为 $(n \pm \Delta \nu_{x,y})f_0$，因此可以得到横向四极振荡频率为 $(n \pm 2\Delta \nu_{x,y})f_0$。

### 2.2 基于条带 BPM 提取横向四极信号

合肥光源储存环上安装了 3 个条带纽扣组合 BPM，其中的条带 BPM 横截面示意图如图 19 所示。可以看到 4 个电极是轴对称的。

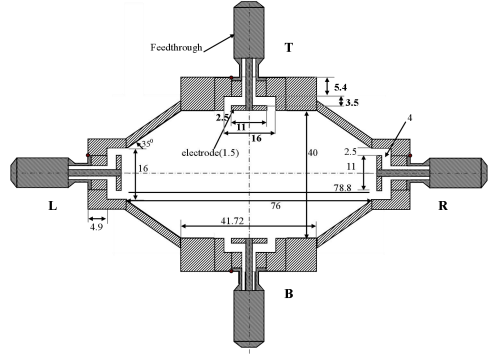

图 19 条带 BPM 横截面示意图

$$\begin{cases} V_R = 1 + Z_{1x} + Z_2 + Z_{3x} + Z_4 + \cdots, \\ V_L = 1 - Z_{1x} + Z_2 - Z_{3x} + Z_4 \cdots, \\ V_T = 1 + Z_{1y} - Z_2 - Z_{3y} + Z_4 + \cdots, \\ V_B = 1 - Z_{1y} - Z_2 + Z_{3y} + Z_4 \cdots. \end{cases} \quad (1)$$

$$\begin{cases} Z_{1x} = 2\dfrac{\sin(\phi/2)}{\phi/2}\dfrac{x_0}{b}, Z_{1y} = 2\dfrac{\sin(\phi/2)}{\phi/2}\dfrac{y_0}{b}, \\ Z_2 = 2\dfrac{\sin\phi}{\phi}\dfrac{\sigma_x^2 - \sigma_y^2 + x_0^2 - y_0^2}{b^2}, \\ Z_{3x} = 2\dfrac{\sin(3\phi/2)}{3\phi/2}\dfrac{3\sigma_x^2 - 3\sigma_y^2 + x_0^2 - 3y_0^2}{b^2}\dfrac{x_0}{b}, \\ Z_{3y} = 2\dfrac{\sin(3\phi/2)}{3\phi/2}\dfrac{3\sigma_x^2 - 3\sigma_y^2 + 3x_0^2 - y_0^2}{b^2}\dfrac{y_0}{b}, \\ Z_4 = \dfrac{\sin(2\phi)}{\phi}\dfrac{3\left(\sigma_x^2 - \sigma_y^2 + x_0^2 - y_0^2\right)^2 - 2\left(x_0^4 + y_0^4\right)}{b^4}. \end{cases} \quad (2)$$

其中 $V_R$, $V_L$, $V_T$, $V_B$ 分别是条带 BPM 的右电极、左电极、上电极和下电极的感应电压信号，当 BPM 的四个电极正交对称时，满足上面公式 1、2。束流的横向四极信号可以通过如下公式（差比和法）计算得到：

$$Q_{\Delta/\Sigma} = \frac{V_R + V_L - V_T - V_B}{V_R + V_L + V_T + V_B} = Z_2 + O\left(\frac{1}{b^6}\right) = 2\frac{\sin\phi}{\phi}\left(\frac{\sigma_x^2 - \sigma_y^2}{b^2} + \frac{x_0^2 - y_0^2}{b^2}\right) + O\left(\frac{1}{b^6}\right) \quad (3)$$

$Q_{\Delta/\Sigma}$ 为计算得到的束流横向四极信号，即对角电极信号求和后差比和，从公式 1、2 中可看出对角电极求和后差比和时，只有 $Z_2$ 会被放大，其他项都会被抵消。

对于非正交对称的情况，也可以通过上述对角电极信号求和后差比和来计算四极信号，只是由于两个方向电极等效的张角 $\phi$ 和到中心的 $b$ 不同而使除 $Z_2$ 项外不能被正好完全抵消，束流的横向四极信号与束流位置 $(x_0, y_0)$ 以及横向尺寸 $(\sigma_x, \sigma_y)$ 的关系可以表示为

$$Q_{\Delta/\Sigma} = Q_0 + S_Q(x_0^2 - y_0^2 + \sigma_x^2 - \sigma_y^2) \quad (4)$$

其中 $Q_0$ 是由于条带 BPM 的 4 个电极之间不是正交对称（只是轴对称，水平电极间距与垂直电极间距不一样）产生的一个直流信号。$S_Q$ 是基于条带 BPM 计算横向四极信号的灵敏度

大小，$(x_0, y_0)$为束流横向位置，$(\sigma_x, \sigma_y)$ 为束流横向位尺寸大小。对于正交对称转 45 度的斜 45 度对称的 BPM，也可以这样计算四极信号，只是公式中的$(x_0, y_0)$和$(\sigma_x, \sigma_y)$应该是相对于旋转 45 度后的坐标系的坐标和尺寸，如果束团有很强的横向四极振荡，$Q_{\Delta/\Sigma}$的频谱也会在$\left(n \pm 2\Delta\nu_{x,y}\right)f_0$这些频点附近看到谱峰，在$f_0$范围内，对应合肥光源的工作点，在 515kHz（$\sigma_x$）、1.188MHz（$\sigma_y$）、3.346MHz（$\sigma_y$）、4.019MHz（$\sigma_x$）这几个频点左右。

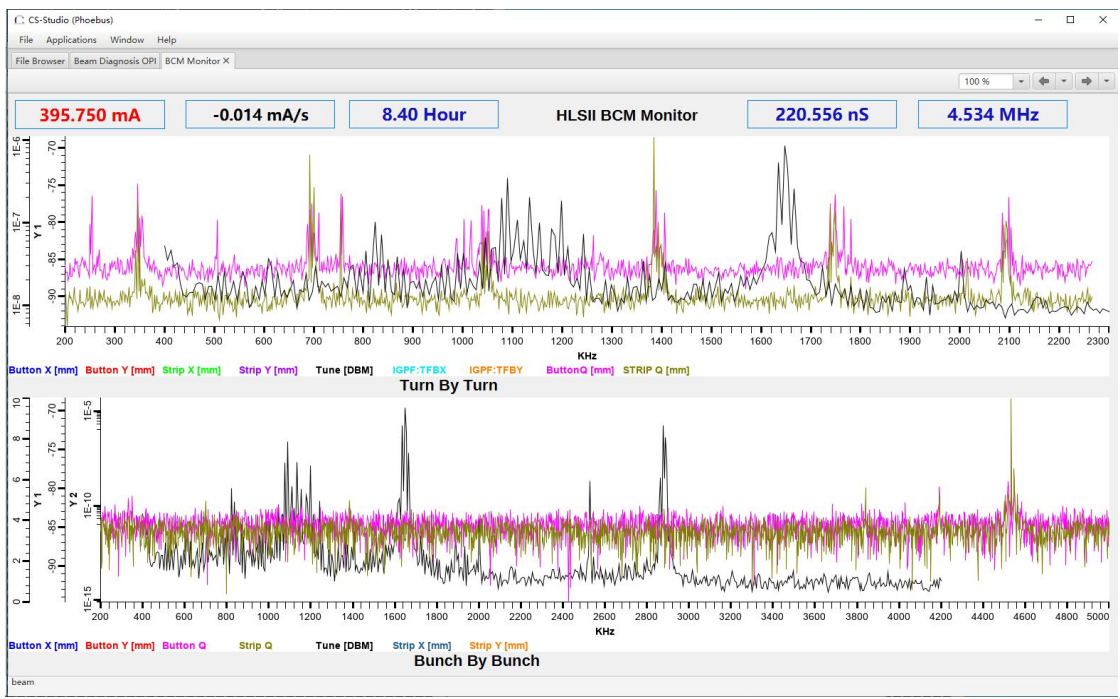

图 20  为运行常态时的 $Q_{\Delta/\Sigma}$ 的谱线

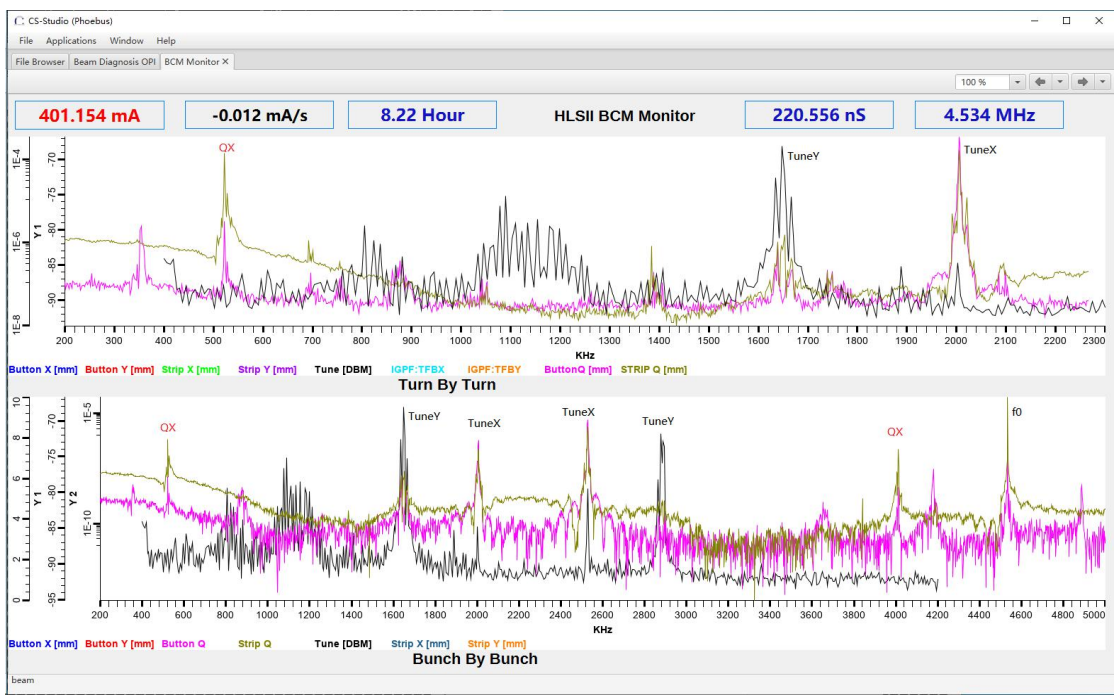

图 21   注入时的 $Q_{\Delta/\Sigma}$ 的谱线

图 20 为运行常态时的 $Q_{\Delta\Sigma}$ 的谱线，粉色曲线为纽扣 BPM，橄榄色为条带 BPM。除去噪声信号计算产生的谱峰，在常态下看不到四极振荡谱峰。图 21 为注入时的 $Q_{\Delta\Sigma}$ 的谱线，合肥光源是水平方向注入，注入时会产生水平方向四极振荡，图中标记 QX 的位置为$\sigma_x$变化对应的水平方向四极振荡谱峰，Turn By Turn 的频谱图只能看到下边带部分，Bunch By Bunch 的频谱图对称的上下边带的四极振荡的谱峰都可以看到。条带的 $Q_{\Delta\Sigma}$ 四极振荡的幅度和工作点处的幅度同量级，在线性标尺下就能够看到，但是纽扣的差两个量级左右，用对数标尺才能看出来，此两幅图条带和纽扣的纵坐标标尺都是对数标尺。

**总结与展望**

此逐束团质心 3 维测量系统已在合肥光源在线工作，基于其处理产生的数据在运行常态时可以得到每个束团的 3 维（横向和纵向）工作点，通过对 BPM 信号的分析，可以得到注入时束团横向尺寸四极振荡的谱峰，这个系统产生的数据，相信通过进一步的分析可以得到更多的结果。

虽然能看到注入时水平方向四极振荡的谱峰，但是更细节的尺寸变化规律还是需要以后搭建逐束团 6 维（质心+尺寸）测量系统做进一步观测。搭建起系统只是开始，进一步的数据分析和信息挖掘需要更多感兴趣的人参与进来。

开发的程序现在的 1.0.X 版本能方便的设置填充模式等参数以通用于各光源，而且对基于示波器即将搭建的逐束团 6 维（质心+尺寸）测量系统，稍作修改和补充即可使用，现已开源在 gitee[9] 上，欢迎兄弟单位使用。

值得一提的是在这半年的系统从零开始搭建到现在的在线运行，在国产 Zstack 虚拟架构下运行的虚拟机，网络和信号处理一直高负荷工作，从来没有过死机，一直稳定运行；国产鼎阳示波器也从来没有过宕机，没有过意外中断连接的情况发生，都是因为改进和调试程序而人为的停止。国产的软件架构和仪器已完全能满足大科学工程高要求的常年在线工作场合的应用。

**参考文献：**